\documentstyle[epsf,epsfig,aps]{revtex}
\draft
\input epsf
\begin{document}
\twocolumn[\hsize\textwidth\columnwidth\hsize\csname@twocolumnfalse%
\endcsname
\title{Absence of the Transition into Abrikosov Vortex State of Two-Dimensional Type-II Superconductor with Weak Pinning.}
\author{A. \ V. \ Nikulov, \ D. \ Yu. \ Remisov}

\address{Institute of Microelectronics Technology and High Purity Materials, Russian Academy of Sciences, 142432 Chernogolovka, Moscow District, RUSSIA}

\author{V. \ A. \ Oboznov}

\address{Institute of Solid State Physics, Russian Academy of Sciences, 142432
Chernogolovka, Moscow District, Russian Federation.}
\maketitle

\begin{abstract}
{The resistive properties of thin amorphous NbO$_{x}$ films
above and below $H_{c2}$ were investigated experimentally. It was shown that
near $H_{c2}$ the excess conductivity dependencies agree with the
predictions of the paraconductivity theory. The current-voltage
characteristics in a perpendicular magnetic field remain Ohmic down to
very low magnetic field. This is interpreted as the absence of the
transition into Abrikosov vortex state of two-dimensional type-II
superconductor with weak pinning.}  \end{abstract}
  ]

 \narrowtext

Before the discovery of high-$T_{c}$ superconductors (HTSC's) the
point of view prevailed that between the first, $H_{c1}$, and the second,
$H_{c2}$, critical fields type II superconductors can be only in
the vortex-lattice state, first described by Abrikosov \cite{abr}.
The unusual properties of HTSC's in a magnetic field have stimulated
considerable interest to fluctuation effects in type-II superconductors.
In addition to the vortex-lattice the existence of other states (vortex glass,
vortex fluid) is assumed in many works \cite{fish,larkin}. In the majority
of papers the influence of fluctuation on mixed state properties of HTSC's was
investigated. But a HTSC is unconvenient object for initial investigation of
this phenomenon. The HTSC's have some peculiarities that are
not quite understood. Besides, all known HTSC samples have a
strong vortex pinning. In most of the works the fluctuation
influence on properties connected with pinning is investigated.
But until now the pinning effect is not described
quantitatively for a real case. A limited number of theories
of weak pinning \cite{larc} can be compared with experimental data
only qualitatively. For this reason numerous experimental
confirmations of fluctuation theories of the HTSC mixed state
are not looked reliable. The experimental data have alternative
explanations almost in all cases.

Therefore we think that investigation of fluctuation
effects in conventional, low-$T_{c}$ type-II superconductors
(LTSC's) with weak pinning is very actual. The investigation
of this objects can help to verify the fluctuation theories
of mixed state. The point of view is widespread that in LTSC
the fluctuation effects are small \cite{fish}. But this opinion is
not correct for the high magnetic field region. Calculation \cite{nik93}
and investigation \cite{nik81} have shown that fluctuation
effects in the mixed state of conventional "dirty" type-II
superconductors are great, no more smaller than in HTSC.

The thermodynamic average order parameter is distributed inhomogeneously
in a superconductor which is in the Abrikosov state. Therefore
properties of type-II superconductors in the mixed state may be separated
on two type.  It is obvious that the magnetization and the specific heat
are connected with the spatial average of the order parameter, whereas the
vortex pinning is connected with inhomogeneous distribution of the order
parameter in the space. It was shown in \cite{tesan,newnik} that the
fluctuation essentially alters the spatial average value of order parameter
inside the critical region only. The experimental dependencies of the
specific heat \cite{thou} and the magnetization \cite{nik84} in the
$H_{c2}$ critical region are described by the fluctuation theory of D-2
dimensional superconductors. The dependencies for bulk superconductor are
described by the theory of one-dimensional superconductor. The resistive
transition in parallel magnetic field of bulk superconductors is also described
by the theory of one-dimensional superconductor \cite{nik93}.

Therefore the fluctuation influence on the order parameter distribution is
the most interesting problem now. There are two different theoretical
approach to this problem. In most works (see for example
\cite{fish}) the fluctuations are considered as a oscillation of the
Abrikosov vortex lattice.  The concept of "vortex lattice melting" has
appeared in this approach. In other works (see for example
\cite{tesan,nik90,newnik}) a revision of the Abrikosov solution is made.
Experimental investigation of this problem is connected with investigation
of vortex pinning because the pinning appearance is connected with
appearance of the inhomogeneous distribution of the order parameter.

It is well known that the vortex pinning causes the non-Ohmic
current-voltage characteristics in perpendicular magnetic field. According
to the classical work \cite{kim}, the linear part of the current-voltage
characteristic at a large current can be described as $V
= R_{f}(I - I_{c})$. $R_{f}$ is the flux flow resistivity. $I_{c}$ is the
dynamic critical current determined by pinning. The experimental
investigations show that the current-voltage characteristics become
non-Ohmic ($I_{c}$ become nonzero) below $H_{c2}$ both in bulk
superconductors \cite{nik81,nik84,nik93,kwok} and in thin films
\cite{kes}. In Refs. \cite{kwok,kes} this qualitative change is
interpreted as a vortex lattice melting whereas in \cite{nik81a,nik93} it
is interpreted as a transition from the normal state into the Abrikosov vortex
state. But the vortex liquid does not represent a new genuine
thermodynamic phase different from the normal state \cite{larkin}.
Therefore these two interpretation coincide.

The position and the width of this transition in bulk superconductor differ
from those in thin film. In bulk conventional superconductors the
current-voltage characteristics become non-Ohmic at some percent below
$H_{c2}$ only \cite{nik81,nik84} whereas in thin films the position of this
transition ("melting") depends on the film thickness \cite{kes} and may be
below $0.5H_{c2}$ \cite{kes}. This difference agrees with the difference of
influence of fluctuation on three- and two-dimensional superconductors
with real size which is predicted by theory \cite{maki,moore,ikeda}.

It was shown in \cite{nik81a} that in bulk superconductor with weak pinning the
width of the transition connected with pinning appearance ("melting"
transition) is very small, much smaller than the width of the specific heat
transition \cite{thou}, the magnetization transition \cite{nik84} and
resistive transition in parallel magnetic field \cite{nik93}, which are
connected with the change of the spatial average value of the order
parameter.  Therefore in \cite{nik90} the transition connected with the
appearance of non-Ohmic current-voltage characteristics (with pinning
appearance) was called a narrow transition, whereas the transition
connected with the change of the spatial average value of the order
parameter was called wide transition. The intrinsic width of the narrow
transition has not been determined. It was determined only in
\cite{nik81a} that the narrow transition of the most homogeneous sample is more
than 10 times narrower than intrinsic wide transition. It was shown in
\cite{nik81} also that not only does the pinning appear (the $I_{c}$ value
becomes non-zero) but also the flux flow resistivity $R_{f}$ decreases
sharply at the narrow transition of bulk superconductors (see \cite{nik93}
also). The resistivity dependencies of thin films do not have sharp feature
and are smooth at the pinning appearance (at "melting") \cite{kes}.

In the present work we investigated thin films. Thin films were studied
before in some works \cite{kes,kapit}. But our investigations have shown
that the position of the "melting" transition is not universal for different
films and depends on a amount of pinning centers in them.  Therefore we
wanted to produce films with a smallest amount of pinning centers. The
results of the investigation of these films are presented here. Following 
\cite{kes} and \cite{nik81a} we will determine the transition (melting
\cite{kes} or transition into Abrikosov state \cite{nik81a}) position as the
point at which the current-voltage characteristics in a perpendicular
magnetic field change qualitatively (become Ohmic on increasing of
magnetic field and become non-ohmic on decreasing one). It is obvious that
this change of current-voltage characteristics is caused by the pinning
appearance (disappearance). Therefore we will connect this transition with
the pinning ($I_{c}$) appearance (disappearance) also.

The Nb, NbN, PbBi, Sn, NbO$_{x}$ films produced by magnetron 
sputtering, pulse laser deposition
and electron beam evaporation were examined. All
films, except some amorphous NbO$_{x}$ film, did not have enough
small vortex pinning and therefore are not suitable for our purpose. For
this reason we mainly studied the NbO$_{x}$ films.

The NbO$_{x}$ films were produced by magnetron sputtering of
Nb in an atmosphere of argon and oxygen. The critical temperature
 $T_{c}$ of films used is equal to 2.37 K and \(dH_{c2}/dT = - 22\)
kOe/K. The film thickness $d = 20$ nm. The normal resistivity
\(\rho_{n}=99 \Omega/\Box \). The temperature dependence of normal
resistivity is very weak. \(1/\rho_{n} \left|d\rho_{n}/dT \right|
< 0.0002 \) in the region 20-40 K, where superconducting fluctuation is
small. The resistivity increases with decreasing temperature. This change
can be connected with weak localization. A magnetic field up to 50 kOe
produced by a superconducting solenoid was measured with relative error
0.0005. The resistivity was measured in perpendicular magnetic field,
with a relative error 0.0001. The 0.01 error in the measurement of the
specific resistivity was due to the inaccurate determination of the
geometric dimensions of the film structure. The temperature was measured
with a relative error 0.001.

\begin{figure}[bhb] \vspace{0.1cm}\hspace{-1.5cm}
\vbox{\hfil\epsfig{figure= 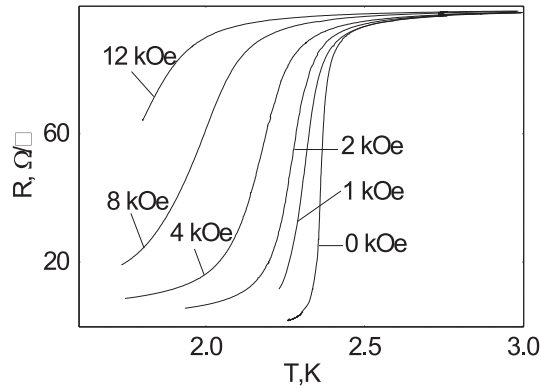, width=7cm,angle=0}\hfil}
\vspace{0.75cm} \caption{Resistive transitions of amorphous NbO$_{x}$ film in different
perpendicular magnetic fields. The film thickness $d = 20$ nm.} \label{fig-1} \end{figure}

The measurement shown that the resistive transition of
NbO$_{x}$ films broadens in a magnetic film as well as of the
HTSC resistive transition (Fig.1). The paraconductivity \(\Delta\sigma =
\sigma - \sigma_{n} \) dependencies above $H_{c2}$ in the
linear approximation region are well described by Ami-Maki theory
\cite{ami78} adapted to a two-dimensional superconductors (Fig.2). The
Maki-Thompson contributions are partly suppressed. The normal conductivity
value \(\sigma_{n}\) was determined from extrapolation of its high temperature
dependence and it is not fit parameter. Therefore the single fit
parameter is a $H_{c2}$ value. The temperature dependence of the fit
$H_{c2}$ values agree with Maki theoretical dependence \cite{maki64}.
Consequently we have one fit parameter, $H_{c2}$(T=0), for all
paraconductivity dependencies. The discrepancy between experimental and
Ami-Maki dependencies near T$_{c2}$ is connected with invalidity of the linear
approximation in the critical region. The calculation of the fluctuation
interaction in Hartree approximation removes this discrepancy (Fig.2).
For this calculation the Ginsburg number, \(D = 2\pi k_{B} T_{c}
/H_{c}^{2}(0)d\xi^{2}(0)\), was used as fit parameter. The only parameter,
which was not determined irrespective, is a thermodynamic critical field
\(H_{c}(0) = -T_{c} (dH_{c} /dT)_{T=T_{c}}\). The fit value of
\(-(dH_{c} /dT)_{T=T_{c}} = 300\) Oe/K. It is not far from this
value for pure Nb, which is equal 472 Oe/K.

\begin{figure}[bhb] \vspace{0.1cm}\hspace{-1.5cm}
\vbox{\hfil\epsfig{figure= 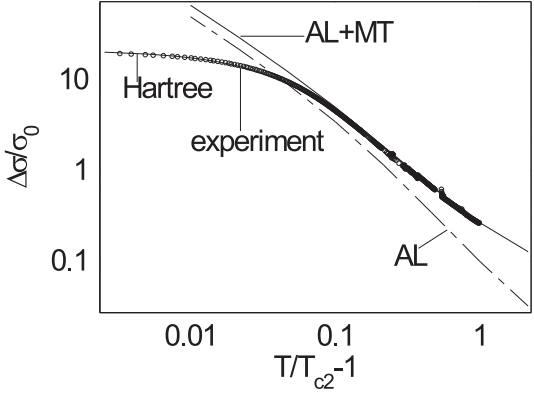,width=7cm,angle=0}\hfil}
\vspace{0.75cm} \caption{Paraconductivity \(\Delta \sigma/\sigma_{0}\) dependencies on
\(T/T_{c2} - 1\) for NbO$_{x}$ film with $d = 20$ nm in perpendicular
magnetic field 12 kOe. The theoretical dependencies for the
Aslamasov-Larkin contribution (AL), for the sum of the Aslamasov-Larkin and
Maki-Thompson contributions (AL+MT) (linear approximation) and for the
Hartree approximation are shown. \(\sigma_{0} = e^{2} /\hbar \). The
Ginsburg number $D = 0.003$.} \label{fig-1} \end{figure}

The paraconductivity investigations show that the
amorphous NbO$_{x}$ films studied are conventional homogeneous
two-dimensional type-II superconductor. But the narrow transition,
which was observed near $H_{c2}$ in bulk type-II
superconductors \cite{nik93,nik81a}, or the "melting" transition, which was
observed in films \cite{kes,kapit}, are not observed in these
films. The current-voltage characteristics remain Ohmic down to very low
magnetic field (Fig.3). The resistivity value decreases gradually with
decreasing magnetic field value (Fig.3).  At low magnetic field it is close
to a flux flow resistivity value (Fig.3) \cite{gorkov}. The resistivity of
a 10 $\mu$m width strip at T = 1.6 K is equal zero up to current value of 10
mA in zero magnetic field and is not equal to zero already at current
value 1 nA in a low magnetic field 100 Oe.

The absence of a nondissipative current, $I_{c}$ = 0, and a
resistivity value which is close to a flux flow resistivity value,
$R_{f}/R_{n} = 0.25H/H_{c2}$, in a low magnetic field are obviously connected
with the absence of vortex pinning. It should be noted that the
absence of nondissipative current can be connected not only with the 
absence of pinning but also with flux creep, particularly vortex lattice moving and so
on. But in these cases the resistivity value differs from the flux flow
resistivity value. This situation was observed earlier in \cite{parks}
where the resistivity is more than 3 orders of magnitude less than flux flow resistivity,
and in other works. Therefore we may say that the pinning absence down to
a magnetic field much lower than $H_{c2}$ is observed first in our work.

As was written above, in thin films the pinning disappears ("melting"
transition occurs) not near $H_{c2}$ (as takes place in bulk
superconductors \cite{nik81,nik84}) but markedly below $H_{c2}$ (at $H =
0.3H_{c2}$ for the film thickness 18 nm and $T/T_{c} = 0.67$ \cite{kes}).
Our investigations show that there are films in which the pinning
does not appear down to H much lower than H at which the "melting" transition
was observed in \cite{kes} (down to $H = 0.1 kOe = 0.006H_{c2}$ for film
thickness 20 nm and $T/T_{c} = 0.67$, see the insertion on Fig.3). This means
that the "melting" transition in our films can occure below $0.006H_{c2}$
(at $T/T_{c} = 0.67$) only. Consequently the position of the pinning
appearance (disappearance) is not universal for different films. Therefore
the theory of vortex lattice melting, used in \cite{kes,kapit}, is not valid
there. The "melting" position depends on a amount of pinning centers. The
possible influence of pinning centers on the fluctuation value was mentioned
in \cite{moore}.

\begin{figure}[bhb] \vspace{0.1cm}\hspace{-1.5cm}
\vbox{\hfil\epsfig{figure= 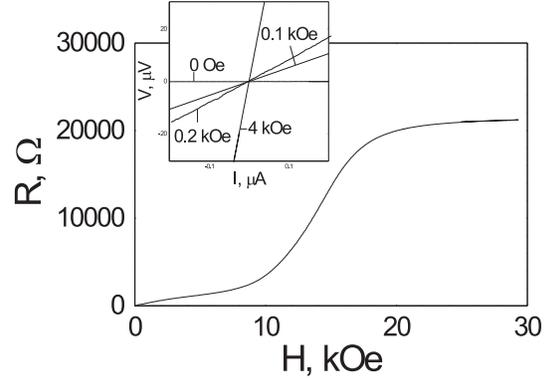,width=7cm,angle=0}\hfil}
\vspace{0.75cm} \caption{The magnetic dependence of the
resistivity of NbO$_{x}$ film structure with sizes: $d = 20$ nm, width =
10 $\mu$m and length = 2250 mm. In the inset the current-voltage curves
in the zero magnetic field and in the different magnetic fields are
shown. $T = 1.6$K. $H_{c2}$ = 16.5 kOe.} \label{fig-1} \end{figure}

In the papers \cite{kes} the resistivity dependencies above "melting" are
compared with flux flow resistivity dependencies obtained in the mean field
approximation. It can not be right because the
fluctuation is big there. It was shown in \cite{nik81a,nik83,nik93} that
the resistivity dependencies of bulk superconductors above "melting" are
described by the paraconductivity theory both above and below $H_{c2}$.
Fig.4 demonstrates that the experimental dependencies of two-dimensional
superconductors can also be described by paraconductivity theory both above
and below $H_{c2}$.  The experimental dependence of \([1+(\Delta \sigma
/\sigma_{n})\sqrt{h/t}]^{-1}\) is a universal function of \( (t-t_{c2}
)/\sqrt{ht}\) (Fig.4), where \(t = T/T_{c}\); \(t_{c2} = T_{c2}/T_{c}\); \(h =
H/H_{c2}(T=0)\); $T_{c2}$ is second critical temperature. This scaling low
follows from the fluctuation theory \cite{newnik}. The universal
experimental dependence is close to the theoretical paraconductivity
dependence obtained in Hartree approximation (Fig.4).

Thus the experimental dependencies are described by the same
paraconductivity dependence both above and below $H_{c2}$. This confirms the
opinion \cite{nik90,ikeda,newnik} that the second critical field line
$H_{c2}(T)$  marks only a crossover from the normal state to a strongly
fluctuating superconducting state with no real phase transition
\cite{larkin}. As was written above, the real phase transition in
the Abrikosov vortex lattice state (melting transition) is connected now
with qualitative changes of the resistive properties in a perpendicular
magnetic field \cite{nik81a,kes,kapit,kwok,nik93}. Therefore the absence of
the qualitative changes of the resistive properties (the current-voltage
characteristics remain Ohmic, $V = R_{f}I$, down to $0.006H_{c2}$, see the
inset in Fig.3) may be interpreted as the absence of the transition
into the Abrikosov vortex lattice state of two-dimensional
superconductors with small amount of pinning centers down to magnetic field
hundreds times smaller than the second critical field.

\begin{figure}[bhb] \vspace{0.1cm}\hspace{-1.5cm}
\vbox{\hfil\epsfig{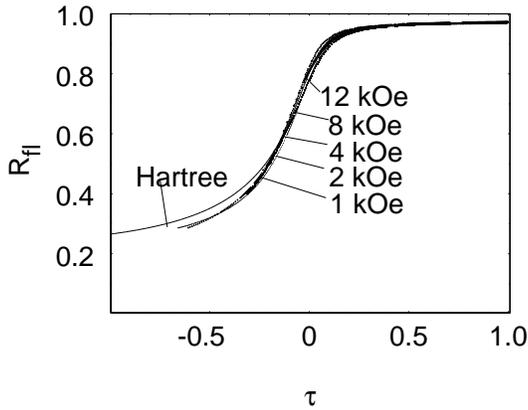}\hfil}
\vspace{0.75cm} \caption{The \(R_{fl} = [1+(\Delta\sigma /\sigma_{N}\sqrt{ht} )]^{-1} \) versus
\(\tau = (t-t_{c2} )/\sqrt{ht}\) dependencies in different magnetic fields. The
line denotes the theoretical dependencies of paraconductivity obtained by
the calculation of the fluctuation interaction in the Hartree
approximation.} \label{fig-1} \end{figure}

This work was supported by the National Scientific Council
on High Temperature Superconductivity, Project No. 92124.
A.V.N. thanks the International Science Foundation for financial
support.


\begin{references}

\bibitem[1]{abr}A.A. Abrikosov, Zh. Eksp. Teor. Fiz. {\bf32}, 1442 (1957)
(Sov.Phys.JETP {\bf5}, 1174 (1957)).

\bibitem[2]{fish}D.S.Fisher, M.P.A. Fisher and D.A. Huse, Phys. Rev. B
{\bf43}, 130 (1991).

\bibitem[3]{larkin} G.Blatter, M.V.Feigel'man, V.B.Geshkenbein, A.I.Larkin,
and V.M.Vinokur, Rev.Mod.Phys. {\bf66}, 1130 (1994).

\bibitem[4]{larc}A.I.  Larkin and Yu.N.  Ovchinikov, J.  Low
Temp.  Phys.  {\bf34}, 409 (1979).

\bibitem[5]{nik93}V.A. Marchenko and A.V.  Nikulov, Physica C {\bf210}, 466
(1993).

\bibitem[6]{nik81}V.A.
Marchenko and A.V. Nikulov, Zh. Eksp.  Teor. Fiz. {\bf80}, 745 (1981) (Sov.
Phys.-JETP {\bf53}, 377 (1981)).

\bibitem[7]{tesan}Zlatko Tesanovich and A.V. Andreev, Phys. Rev. B
{\bf49}, 4064 (1994). Zlatko Tesanovich, Lei Xing, Lev Bulaevskii,
Quiang Li, and M. Suenada, Phys. Rev. Lett. {\bf69}, 3563 (1992). Zlatko
Tesanovich, Physica C {\bf220}, 303, (1994).

\bibitem[8]{newnik}Nikulov A.V., Physica C {\bf235-240}, 1945
(1994); Nikulov A.V. submitted to Phys. Rev. B

\bibitem[9]{thou}D.J. Thouless, Phys. Rev.
Lett.  {\bf34}, 946 (1975); S.P. Furrent and C.E. Gough, Phys. Rev. Lett.
{\bf34}, 943 (1975)

\bibitem[10]{nik84}V.A. Marchenko and A.V. Nikulov, Zh.
Eksp. Teor. Fiz. {\bf86}, 1395 (1984) (Sov.Phys.-JETP {\bf59}, 815 (1984)).

\bibitem[11]{nik90}A.V. Nikulov, Supercond. Sci. Technol. {\bf3}, 377
(1990).

\bibitem[12]{kim} Y.B.Kim, C.F.Hemstead, and A.R.Strand, Phys.Rev.A
{\bf136}, 1163 (1965).

\bibitem[13]{kwok} W.K.Kwok, U.Welp, G.W.Crabtree, K.G.Vandervoort,
R.Hulscher, and J.Lie, Phys.Rev.Lett. {\bf64}, 969 (1990); W.K.Kwok,
J.Fendrich, S.Flesher, U.Welp, J.Downey, and G.W.Crabtree, Phys.Rev.Lett.
{\bf72}, 1092 (1994).

\bibitem[14]{kes} P.Koorevar, P.H.Kes, A.E.Koshelev, and Aarts,
Phys.Rev.Lett. {\bf72}, 3250 (1994); P.Berghuis and P.H.Kes, Phys.Rev. B
{\bf47}, 262 (1993); P.Berghuis, A.L.F.van der Slot, and P.H.Kes,
Phys.Rev.Lett. {\bf65}, 2583 (1990).

\bibitem[15]{nik81a}V.A. Marchenko and A.V. Nikulov, Pisma Zh.
Eksp. Teor. Fiz. {\bf34}, 19 (1981) (JETP Lett. {\bf34}, 17 (1981)).

\bibitem[16]{maki}K. Maki and H. Takayama, Progr.
Theor.  Phys.  {\bf46}, 1651 (1971).

\bibitem[17]{moore}Moore M.A., Phys.Rev.  B{\bf39}, 136 (1989); Moore M.A.,
Phys. Rev. B{\bf45}, 7336 (1992); H.H.Lee and M.A.Moore, Phys.Rev. B
{\bf49}, 9240 (1994).

\bibitem[18]{ikeda}R.Ikeda, T.Ohmi and T.Tsuneto, J. Phys. Soc.Jap.
{\bf61}, 254 (1992).

\bibitem[19]{kapit} A.Yazdani, W.R.White, M.R.Hahn, M.Gabay, M.R.Beasley,
and A.Kapitulnik, Phys.Rev.Lett. {\bf70}, 505 (1993).

\bibitem[20]{ami78}S. Ami and K. Maki, Phys.Rev. B {\bf18}, 4714 (1978).

\bibitem[21]{maki64} K. Maki, Physics {\bf1}, 21 (1964).

\bibitem[22]{gorkov} L.P.Gor'kov and N.B.Kopnin, Sov.Phys.Usp. {\bf18}, 496
(1975)

\bibitem[23] {parks} P.M.Horn and R.D.Parks, Phys.Rev. B {\bf4}, 2178 (1971)

\bibitem[24] {nik83} V.A.Marchenko and A.V.Nikulov, Fiz.Nizk.Temp. {\bf9},
816 (1983).

\end{references}
\end{document}